\documentclass[namedreferences]{solarphysics}
\usepackage[optionalrh]{spr-sola-addons}
\usepackage{epsfig}          
\usepackage{graphicx}        
\usepackage{courier}         
\usepackage{savesym}
\savesymbol{iint}
\savesymbol{iiint}
\usepackage{amsmath}
\usepackage{wasysym}
\usepackage{color}           
\usepackage{url}             
            
\usepackage{amssymb}
\usepackage{soul}
\usepackage{ulem}
\usepackage[bf,small,tableposition=top]{caption}
\usepackage{subfig}
\usepackage{enumerate}

\newcommand{\pmV}{pmV$^{-1}$}
\newcommand{\mAV}{m\text{\AA}V$^{-1}$}
\newcommand{\degC}{$^\circ$C}

\newcommand{\aap}{    {\it Astron. Astroph.}}
\newcommand{\aaps}{   {\it Astron. Astrophys. Suppl.}}

\newcommand{\AO}{     {\it Appl. Opt.}}

\newcommand{\jaa}{    {\it J. Astrophys. Astron.}}
\newcommand{\josa}{   {\it J. Opt. Soc. Am.}}

\newcommand{\solphys}{{\it Solar Phys.}}

\newcommand{\BASI}{   {\it Bull. Astron. Soc. India}}

\begin{document}
\begin{opening}
\title{Narrow-Band Imaging System for the {\it Multi-application Solar Telescope} at Udaipur Solar Observatory: Characterization of Lithium Niobate Etalons}
\author{A.~\surname{Raja~Bayanna}\sep
           \surname{Shibu~K.~Mathew}\sep
         P.~\surname{Venkatakrishnan}\sep
	 N.~\surname{Srivastava}
       }
\runningauthor{A.~Raja Bayanna et al.}
\runningtitle{Narrow band imager for MAST}

\institute{Udaipur Solar Observatory, Physical Research Laboratory,  Dewali, Badi road, Udaipur, India-313004\\
email: \url{bayanna@prl.res.in}\\ 
}
\begin{abstract}
{\it Multi-application Solar Telescope} is a 50 cm off-axis Gregorian telescope that has been installed at the lake site of Udaipur Solar Observatory. For quasi-simultaneous photospheric and chromospheric observations, a narrow-band imager has been developed as one of the back-end instruments for this telescope. Narrow-band imaging is achieved using two lithium niobate Fabry-Perot etalons working in tandem as a filter. This filter can be tuned to different wavelengths by changing either voltage, tilt or temperature of the etalons. To characterize the etalons, a Littrow spectrograph was set up, in conjunction  with a 15 cm Carl Zeiss Coud\'{e} solar telescope. The etalons were calibrated for the solar spectral lines Fe {\sc i} 6173~\text{\AA\/}, and Ca {\sc ii} 8542~\text{\AA\/}. In this work, we discuss the characterization of the Fabry-Perot etalons, specifically the temperature and voltage tuning of the system for the spectral lines proposed for observations. We present the details of the calibration set-up and various tuning parameters. We also present solar images obtained using the system.
\end{abstract}
\keywords{Instrumentation and data management, Integrated Sun Observations, Spectral line}
\end{opening}

\section{Introduction}
\label{intro}
Measuring the magnetic and velocity fields on the Sun is the key to understanding the solar activity and fine structure of sunspots, filaments, {\it etc}. One of the main techniques techniques for measuring solar magnetic and velocity fields involves spectroscopy, which can be carried out using either a high-resolution spectrograph or a tunable narrow-band filter in an imaging system \cite{Zirin}. Imaging spectroscopy allows us to obtain two-dimensional images (with certain field-of-view (FOV)) of the solar surface but requires scanning across multiple wavelength positions to build up the spectra. On the other hand, a slit-based spectrograph obtains the spectra but requires scanning across the FOV spatially to build~up a two-dimensional image of the solar surface \cite{Judge2010}. In both cases, the observations from ground-based telescopes are affected by atmospheric turbulence. A number of {\it post-facto} techniques exist for imaging and imaging spectroscopy to deal with effects of atmospheric turbulence, but these techniques are still evolving for observations from a slit-based spectrograph \cite{Beck2011}.

Conventionally, narrow-band imaging and imaging spectroscopy can be carried out by using wavelength-tunable Lyot filters \cite{Lyot, John, Stix}, which can be slow because of mechanical movements involved in the tuning process. Using fast tunable narrow-band filters, two-dimensional images can be obtained by scanning across the spectral lines in shorter time periods (depending upon the number of wavelengths to be scanned along the line profile). Two-dimensional images, with sparsely sampled spectral information, provide sufficient opportunity to use inversion techniques to retrieve the magnetic and velocity fields \cite{MEinversion, Borrero2011, Valentine2011}. For example, the {\it CRisp Imaging Spectropolarimeter} (CRISP: \inlinecite {CRISPSc2}) installed at the Swedish 1-m telescope records Stokes images at 11 positions along the line profile of Fe~{\sc i} at 6302~\text{\AA\/} in steps of 48~m\text{\AA\/}. The Stokes images were processed with the Milne-Eddington inversion software Helium Line Information Extractor (HELIX) \cite{MEinversion} to obtain the atmospheric parameters. \inlinecite{CRISPSc2} used CRISP data  to study the fine structure of the sunspot penumbra. \inlinecite{wiegelmann} also used the Stokes vector maps of the quiet Sun obtained with the Imaging Magnetograph eXperiment (IMaX) on the {\it Sunrise}~\cite{sunrise} telescope, to study the magnetic connectivity between the layers of the solar atmosphere. IMaX samples the line profile of Fe {\sc I} at 5250.2~\text{\AA\/} at a maximum of 12 positions. In this case,the inversion code Very Fast Inversion of the Stokes Vector (VFISV) developed by \inlinecite{Borrero2011} is used to obtain atmospheric parameters. \inlinecite{wiegelmann} used the data to obtain the 3D structure of magnetic loops using extrapolation methods. Similarly, the {\it Interferometric Bidimensional Spectrometer} (IBIS) \cite{IBIS2006} installed at the {\it Dunn Solar Telescope} at Sacramento Peak (New Mexico, USA) records multiwavelength observations of the solar atmosphere to understand various solar phenomena. It scans the line profiles of Fe~{\sc i}, H${_\alpha}$, and Ca~{\sc ii} at 14, 22, and 21 positions, respectively.

Some of the recent instruments use Fabry-Perot etalons to achieve narrow-band imaging with a fast cadence \cite{Burton1987, Rust,  Shibu1997, Shibu1998, Kentischer1998, Tritschler2002, IBIS2006, SVM2006, Scharmer2006, GPI2008, Kleint2011, Valentine2011}. However, the electro-optically tunable solid-state etalons provide a smaller wavelength shift than air-spaced etalons for the same FOV because of their higher refractive index. This makes solid-state etalons of electro-optical materials such as lithium niobate (LN) useful for a larger FOV \cite{Netterfield, Desai1996, Gosain2002, Valentine2011}. After the successful use in ground-based and balloon-borne experiments, LN etalons have also been proposed for some of the future space experiments such as {\it Solar Orbiter} \cite{Shibu_so, Schuhle, Fahmy}.

\begin{table}
\caption{Specifications of the Fabry-Perot etalons at 6328~\text{\AA} as provided by CSIRO, Australia.}
\begin{tabular}{c|c|c|c|c|c|c|c|c}
\hline
       &          & Clear    & d$^\bot$  & Refractive       & R$^\top$   &FSR$^\dagger$  & FWHM$^\ddagger$ &         \\
       &LiNb$0_3$ & aperture & (mm)      & index at         &   (\%)     &(\text{\AA})   & (m\text{\AA})   & Finesse \\
       &          &          &           & 6328~\text{\AA}  &            &               &                 &         \\
\hline
FP$_1$ &Z-cut     & 60 mm    & 0.226    & 2.286             & 93.30      & 3.6           & $\sim$150       & $\sim$24.0\\
FP$_2$ &Z-cut     & 60 mm    & 0.577    & 2.286             & 93.60      & 1.4           & $\sim$70        & $\sim$20.0 \\
\hline
\multicolumn{9}{c}{ Maximum recommended tuning voltage: $\pm$ 3000 V} \\
\multicolumn{9}{c}{ Maximum recommended rate of change of voltage: $\pm$ 1500 Vs$^{-1}$} \\
\multicolumn{9}{c}{ Maximum recommended operating temperature: 45$^\circ$C} \\
\hline
\end{tabular}
$\bot$ : Thickness, \quad $\top$ : Reflectivity, \quad $\dagger$ : Free spectral range,\quad $\ddagger$ : Full-width at half maximum
\end{table}

At the Udaipur Solar Observatory (USO), a narrow-band imaging system (imager) has been developed using two LN etalons in tandem with an appropriate thickness ratio. Using the etalons in tandem effectively increases the free spectral range (FSR) and the spectral resolution of the system \cite{Kentischer1998, Tritschler2002}. Although inter-reflections between multi-etalons cause ghost images, these can be eliminated or minimized either by tilting the etalons with respect to each other, or by placing a prefilter in between the etalons or by both \cite{Tritschler2002, Kleint2011, Valentine2011}.

The etalons for the imager at USO were procured from the Common Wealth Scientific and Industrial Research Organization (CSIRO), Australia. Table~1 lists their relevant specifications. The imager will be used as a back-end instrument with the {\it Multi-application Solar Telescope} (MAST) (\opencite{AMOS2008}, \citeyear{AMOS2010}, \opencite{ShibuMAST}). MAST is an off-axis Gregorian a-focal telescope with a clear aperture of 50~cm. This telescope was developed by Advanced Mechanical and Optical Systems (AMOS), Belgium, and has been installed at the lake site of USO, India. Along with a polarimeter, the imager will be used for measuring the magnetic and velocity fields of active regions in the photosphere and chromosphere. The initial aim is to carry out quasi-simultaneous observations in the two solar spectral lines Fe {\sc i} at 6173~\text{\AA\/}, and Ca {\sc ii} at 8542~\text{\AA\/}, which are formed in the photosphere and chromosphere, respectively.

In this article, we discuss the calibration of the LN etalons in detail and evaluate their suitability for observations at the two spectral lines mentioned above. In Section~2 we discuss various tuning options for the LN etalons, and in Section~3  the optical set-up used to calibrate of these etalons is described. Results obtained from the voltage and temperature tuning of the system are explained in Section 4. Some of the preliminary images taken with the imager are shown in Section~5.

\section{Lithium Niobate Fabry-Perot Etalon}
Fabry-Perot (FP) etalons produce a transmission spectrum (commonly known as channels) according to the Airy formula \cite{BornWolf},
\begin{eqnarray}
I_t = \frac{1}{\left\{{1+F\;sin^2\:\frac{\delta}{2}}\right\}}\:I_i,
\end{eqnarray}
where $F = {4\:R}/{(1-R)^2}$,~$\delta = \frac{4\:\pi}{\lambda}\:n_o\:d\:\cos\:\theta$,~$R = \sqrt{r_1\:r_2}$. Here, $r_1$ and $r_2$ are the coefficients of the reflectance of the surfaces and $I_i$ and $I_t$ are the incident and transmitted intensities, respectively. $n_o$, $d$, and {\bf $\theta$} are the refractive index of the medium, spacing between the reflective surfaces, and the angle of refraction, respectively.

The working principle of the LN Fabry-Perot etalon is the same as that of any other etalon, except that the (partially) reflecting surfaces consist of highly polished faces of a LN wafer. Thus the crystal itself acts as a medium between the reflecting surfaces. To tune the etalon to different wavelengths, the electro-optic property of the LN substrate is used. Applying a voltage to this substrate in effect changes the refractive index of the crystal and thus produces a shift in the wavelength of the transmission channel \cite{Yariv,Bonaccini,Ghatak}. The etalons used in this work are made of a birefringent, uni-axial z-cut LN crystal. The light propagating along the z-axis experiences a single refractive index ($n_x = n_y = n_o$), which can be altered by the application of an electric field. The change in the refractive index ($\Delta{n_o}$) (neglecting the effect of converse piezoelectric effect) is given by

\begin{eqnarray}
\Delta{n_o} = \frac{1}{2}~{n_o}^3~E_z~r_{13},
\end{eqnarray}
where ${n_o}$ is the refractive index of the ordinary ray at wavelength $\lambda_0$, $E_z$ is the electric field produced by the application of the voltage ($V$), and $r_{13}$ is the relevant un-clamped electro-optic coefficient. The change in wavelength with the application of the voltage is given by,
\begin{eqnarray}
\delta\lambda =\frac{1}{2}~\frac{{n_o}^2~V~r_{13}~{\lambda}_{0}}{d}
\end{eqnarray}

This is the most often used property of the LN crystal for tuning the filter to different wavelengths. The tuning rate is of the order of 200--500~m\text{\AA}~s$^{-1}$ depending on the thickness of the etalon. To tune the etalon for the entire wavelength range within two transmission peaks, the minimum voltage required is half the FSR voltage, which is given  by $\lambda_0/(2~{n_o}^{3}~r_{13})$. Half the FSR voltages estimated for 6173~\text{\AA} and 8542~\text{\AA} are around 5000~V and 8000~V, respectively. This is more than the recommended safe voltage for LN etalons (see Table 1). The other two options for tuning the LN etalons are by varying the temperature and tilt of the etalons. Tilting the etalons changes the peak wavelength position, and for small angles the change in wavelength is given by $\delta\lambda = -\lambda~\theta^{2}/(2~{n_o}^2)$.

In our experiment, the temperature tunability was used to fix the initial position of the transmission channel at the optimum wavelength position. Using voltage tuning, the desired spectral lines were then scanned. We have not explored the tilt tunability of the etalons in the present set-up. Sections~4.2 and 4.3 describe the temperature and voltage tuning of individual etalons, respectively, and Table~3 shows the values obtained after tuning. Section 4.4 describes the tuning of the etalons in tandem.

\begin{figure*}[!b]
\begin{center}
\hspace{0cm}
\includegraphics[width=0.330\textwidth, angle = 90]{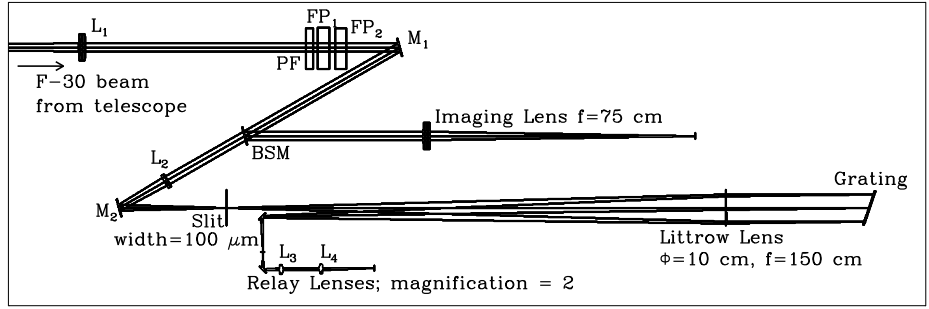}
\caption{Optical layout of the spectrograph and imaging set-up. BSM is the beam-steering mirror for directing the light transmitted from the narrow-band filter (FP$_1$ and FP$_2$) toward the imaging set-up. The prefilter (PF) wheel consists of two prefilters centered on 6173~\text{\AA} and 8542~\text{\AA}. M$_1$ and M$_2$ are folding mirrors.}
\label{SGlayout}
\end{center}
\end{figure*}

\section{Optical set-up for Calibration of Lithium Niobate Etalons and Narrow-band Imaging} 
\label{setup}
For the test and calibration of the LN etalons, we used a Littrow spectrograph set-up as shown in Figure 1, in conjunction with a Coud\'{e} telescope with a 15~cm clear aperture as the light feed. The  F$\#$30 beam (see Figure~1) from the telescope was collimated and reimaged using lenses L$_1$ and L$_2$. Light passing through the slit (of width 100 $\mu$m) was collimated using a lens of focal length 150~cm. The collimated light dispersed by a plane grating (1200 l/mm) was reimaged using the same Littrow lens. The spectrum obtained was magnified by a factor of two at the CCD plane using relay lenses L3 and L4. On the CCD, the dispersion {\it per} pixel (pixel size = 6.45~$\mu$m) is 17.45~m\text{\AA} and 14.85~m\text{\AA} at 6173~\text{\AA} and 8542~\text{\AA}, respectively. The spectral coverage of the spectrograph is 17.8~\text{\AA} and 15.2~\text{\AA} at 6173~\text{\AA} and 8542~\text{\AA}, respectively. The LN etalons (FP$_1$ and FP$_2$) and the prefilter were introduced in the collimated beam after L1. For imaging, a beam-steering mirror (BSM) was inserted in the optical path before L2 to reflect the light toward a lens of focal length 75 cm. This makes a final F$\#$30 beam, with an image scale of  around 0.3 arcsec {\it per} pixel on the CCD camera.

\begin{figure*}[!t]
\begin{center}
\hspace{0cm}
\includegraphics[width=0.235\textwidth, angle = 90]{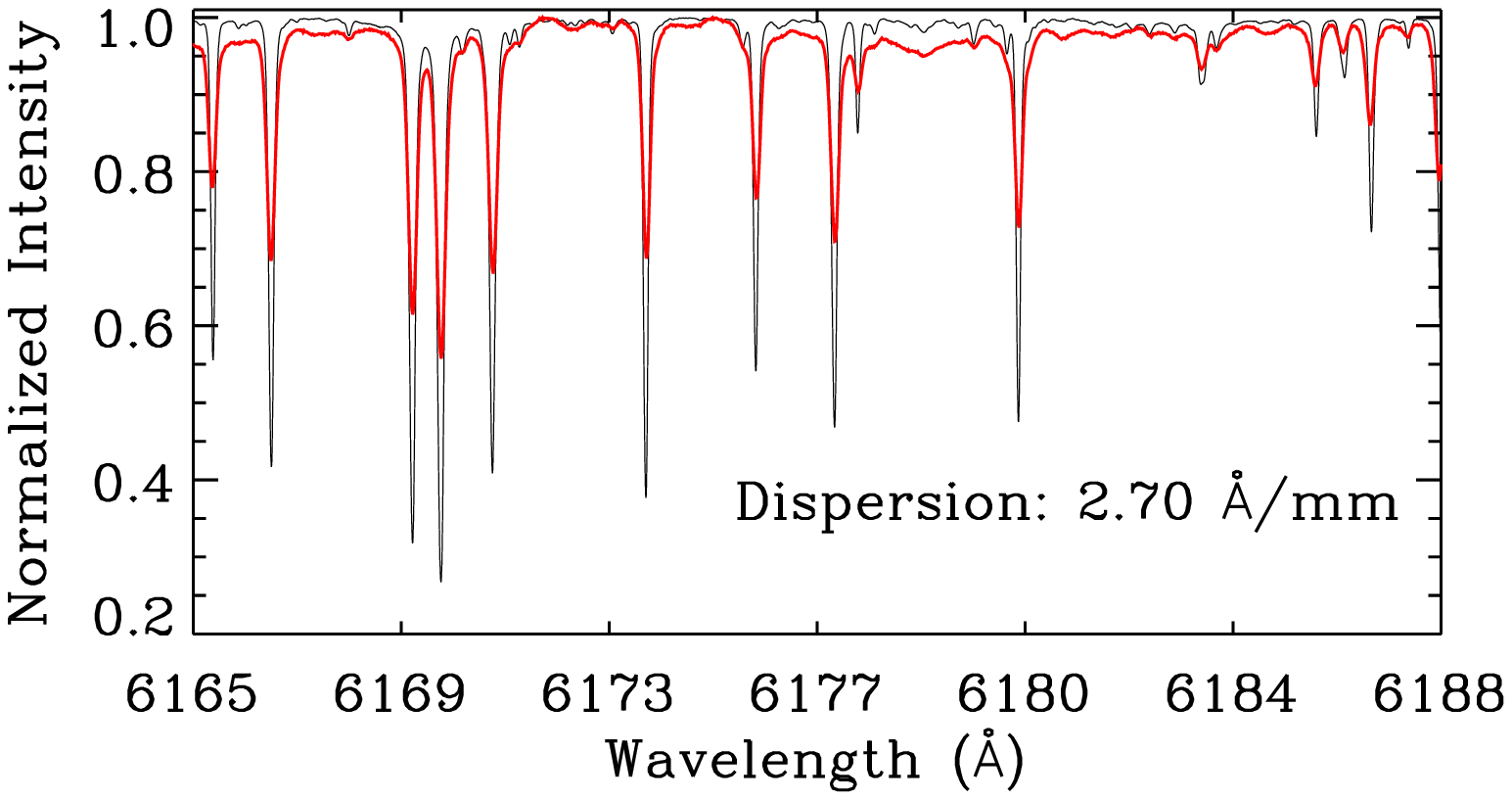}
\hspace{0.2cm}
\includegraphics[width=0.235\textwidth, angle = 90]{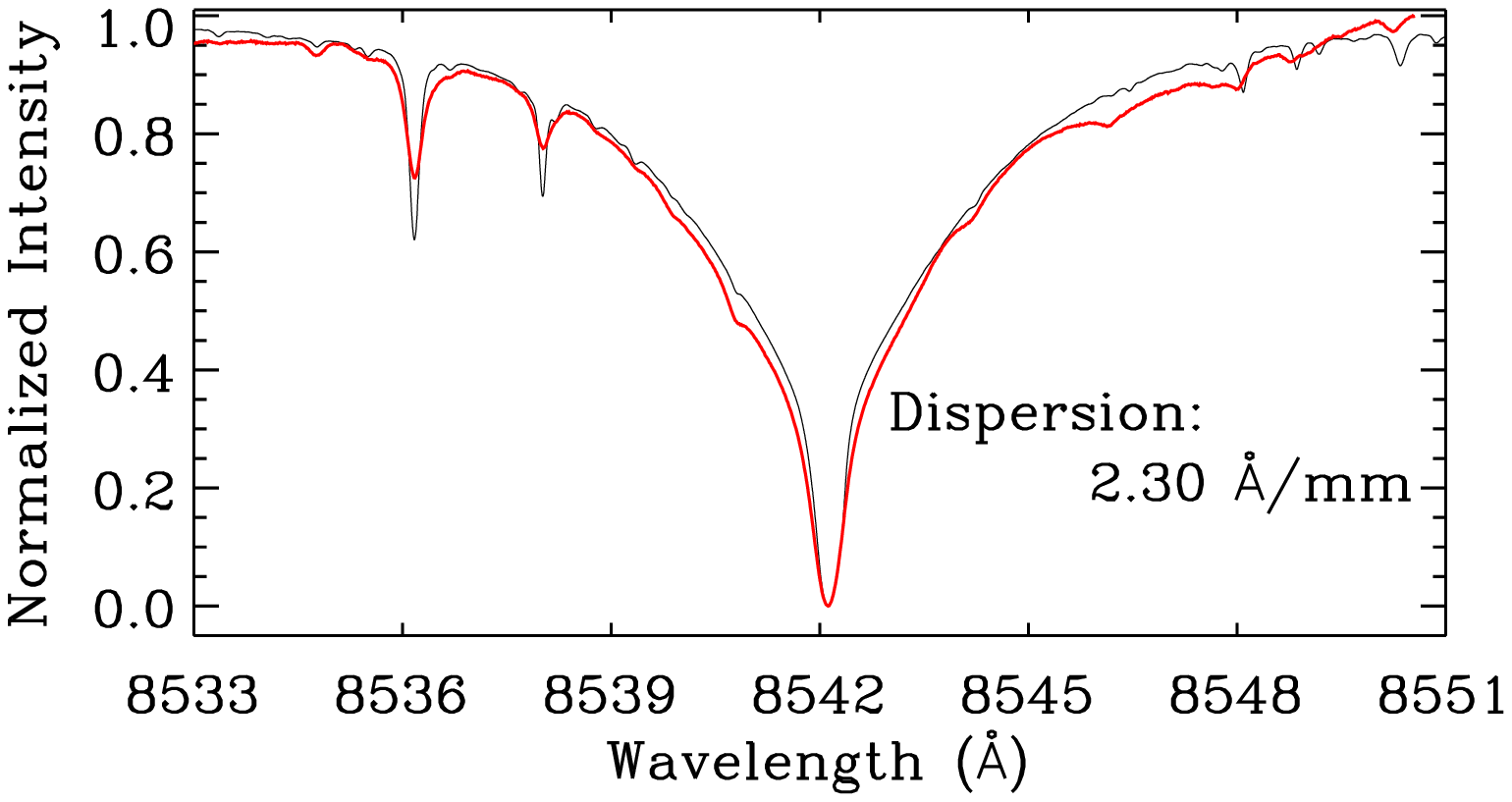}
\caption{Left and right panels show a comparison of the spectrum (red continuous line) obtained using the optical set-up described in Figure 1 with that obtained from the BASS2000 archive (black continuous line) for the 6173~\text{\AA} and 8542~\text{\AA} lines, respectively.}
\label{Spectrum}
\end{center}
\end{figure*}

Two different optical configurations are generally employed for imaging spectroscopy: a) FP etalons in a collimated beam near the re-imaged pupil plane, and b) FP etalons in a telecentric configuration. The merits and demerits of both configurations have been explained by \inlinecite{IBIS2006} and \inlinecite{Scharmer2006}. However, as our primary goal is to characterize the etalons, we preferred a collimated set-up\footnote{Telecentric configuration with F\#$\approx$100 will be chosen in the future for solar observations using this instrument with MAST.}. A maximum blue shift of 11~m\text{\AA} is expected at 6173~\text{\AA} due to the collimated set-up. Figure~\ref{Spectrum} shows the plots of the solar spectrum obtained with our spectrograph set-up and that taken from the BASS2000\footnote{BASS2000: Solar survey archive, $http://bass2000.obspm.fr$, \opencite{BASS2000}} solar survey spectral atlas for comparison.

\begin{figure*}[!b]
\begin{center}
\includegraphics[width=0.198\textwidth, angle = 90]{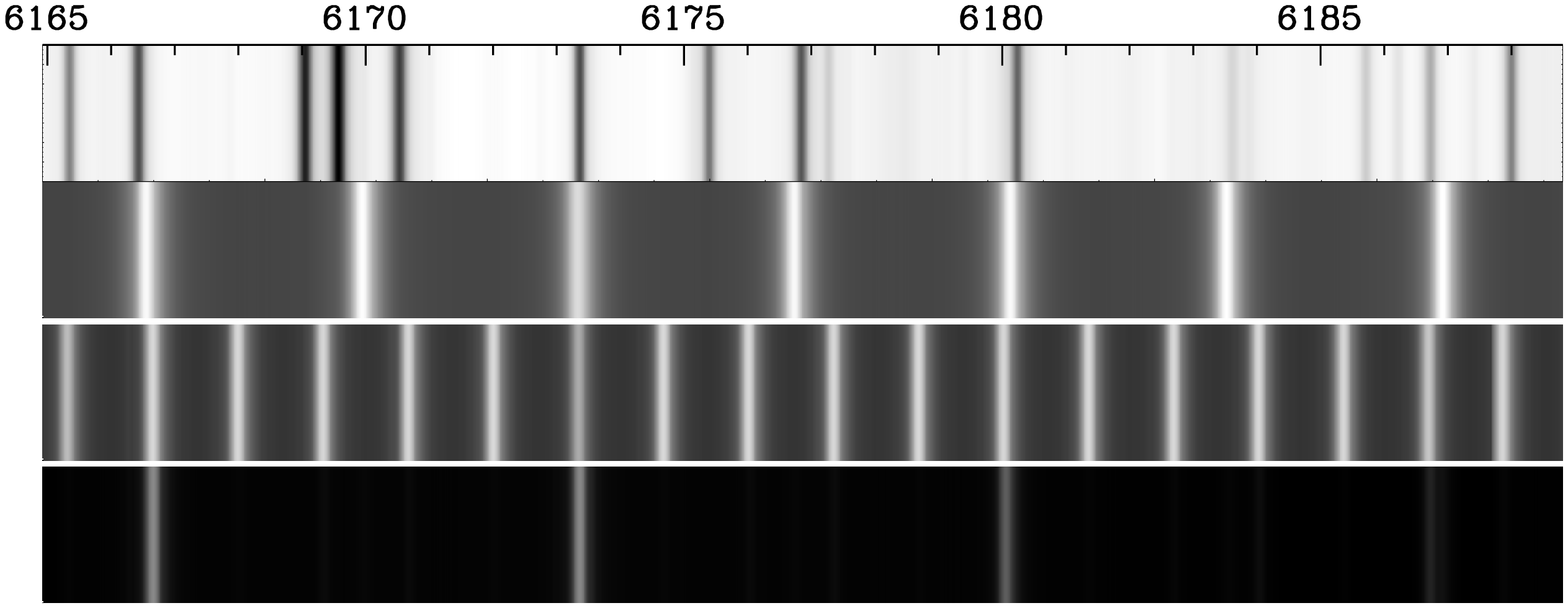}
\includegraphics[width=0.198\textwidth, angle = 90]{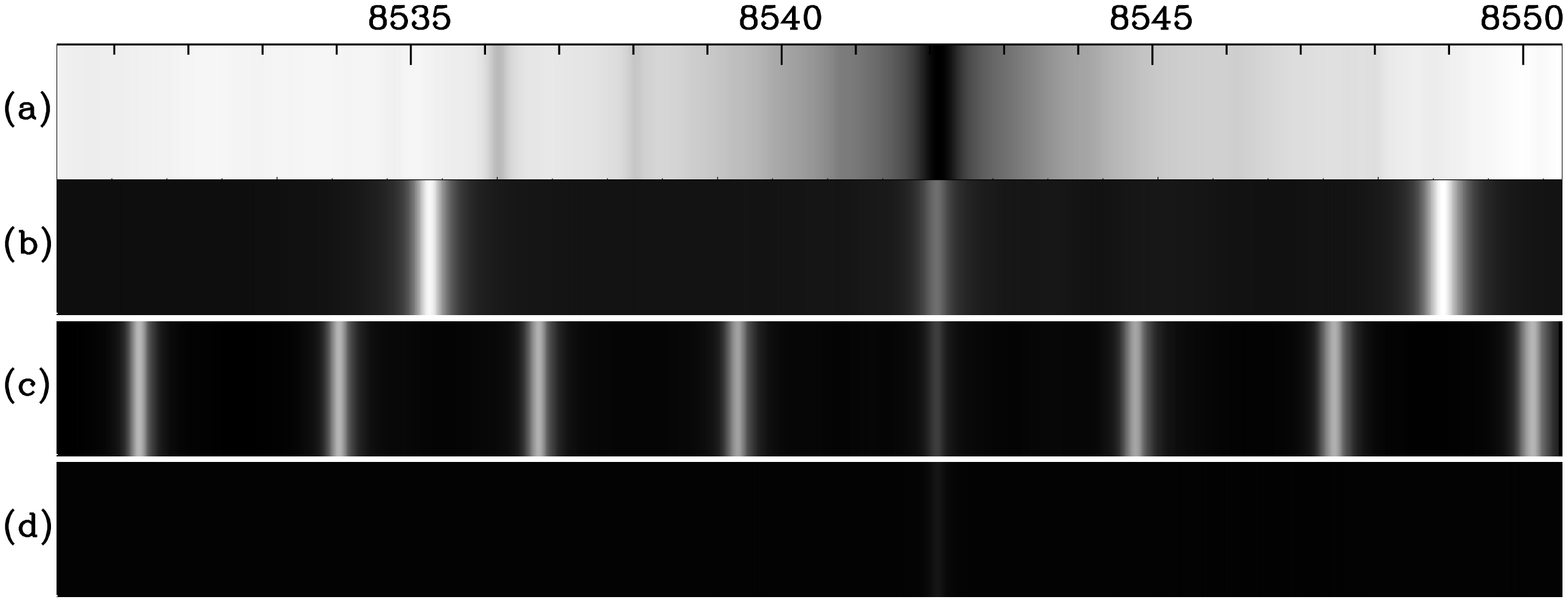}
\caption{ (a) Observed solar spectrum, (b) channel spectrum of FP$_1$, (c) channel spectrum of FP$_2$, (d) channel spectrum of the FPs in tandem near the crossover voltage (the voltage where the channel crosses the line center). Left panels correspond to the 6173~\text{\AA} line, right panels to the 8542~\text{\AA} lines. The intensity in panels (b), (c), and (d) are in logarithmic scale.}
\label{FSRaFWHM}
\end{center}
\end{figure*}

\section{Characterization of the LN Etalons}
\label{charactrization}

Channel spectra of both LN etalons were recorded by placing them in a collimated beam after lens L1. Figure~3 shows the spectra obtained with the optical set-up. The top panel show the solar spectra, which include the 6173~\text{\AA\/} (left) and 8542~\text{\AA\/} (right) spectral lines. The remaining three panels, from top to bottom, show the transmission channels for FP$_1$ (thinner, 226 \text{$\mu$m}), FP$_2$ (thicker, 577 \text{$\mu$m}) and the FPs in tandem in logarithmic scale. It is worth noting that the FSR (wavelength separation between two transmission peaks) increases for the tandem configuration. From these measurements, we computed various parameters for the FPs that are essential for using them in the imager set-up at specific wavelengths. Some of the important estimated  parameters are the FSR, the full width at half maximum (FWHM), and voltage and temperature tuning rates. In the following sections, we discuss the methodology adopted to obtain these parameters.

\subsection{Free Spectral Range and Full Width at Half Maximum}
The recorded channel spectra were analyzed to obtain the FSR and the FWHM for the FPs individually and in tandem. The central 200 pixels of the recorded spectrum corresponding to a FOV of one arcmin were averaged in the slit direction to obtain the transmission profile of the etalons. The wavelength scale was calculated from the solar spectra, which were recorded before placing the etalons in the beam path. We have used almost all the channels in our recorded channel spectrum to compute the mean FSR and FWHM. Table~2 shows the measured values of these parameters.
\begin{table}[!t]
\caption{Measured values of the FSR and FWHM of the FPs}
\begin{tabular}{l|l|l|l|l}
\hline
&Parameter&FP1&FP2&Tandem\\
\hline
~~~~~~~6173~\text{\AA}&&&\\
&FSR~(\text{\AA})   & 3.400 & 1.350 & 6.700 \\
&FWHM~(m\text{\AA}) & 196.8 & 154.0 & 150.3\\
\hline
~~~~~~8542~\text{\AA}&&&\\
&FSR~(\text{\AA})   & 6.830 & 2.690 & >10.0 \\
&FWHM~(m\text{\AA}) & 273.3 & 190.0 & 190.6 \\
\hline
\end{tabular}
\end{table}
Although the FSR obtained nearly matches the values specified by the vendor, we observed a larger FWHM than for the specified values. The difference is mainly attributed to the low spectral resolution due to the larger width of the spectrograph slit. We estimated the FWHM of the spectrograph $(\Delta\lambda_s)$ from the FWHMs of the recorded 6173~\text{\AA} line profile $(\Delta\lambda_r)$ and that obtained from the BASS2000 atlas $(\Delta\lambda_c)$ using the relation $(\Delta\lambda_r)^2$ = $(\Delta\lambda_s)^2$+$(\Delta\lambda_c)^2$. The FWHM of the spectrograph is estimated to be about 140~m\text{\AA}. Thus, the measured FWHMs are (roughly) consistent with the specifications in Table 1, and with the actual resolution of the spectrograph.

\subsection{Temperature tuning of the LN etalons}
The refractive index of the LN is sensitive to temperature. To avoid any wavelength shift during observations, we placed the etalons in temperature-stabilized enclosures. These enclosures keep the etalons within $\pm$0.05\degC~around the set temperature to maintain a wavelength stability of about $\pm$1.2~m\text{\AA}.

The temperature of the enclosure can be set to between 23\degC~and 45\degC. Temperature-sensitivity measurements were carried out by setting the temperature of the enclosure between 38\degC~to 23\degC, in steps of -3\degC. 30-minute interval between each step was allowed for the temperature to stabilize to the set value. At each temperature, voltage was applied to the FPs from -3~kV to +3~kV in steps of 100~V and the corresponding shift in the wavelength of the transmission channel was measured. We also estimated the crossover voltage (CoV), {\it i.e.} the voltage where the transmission channel crosses the center of the solar spectral line.

\begin{figure}
\begin{center}
\includegraphics[width=0.48\textwidth]{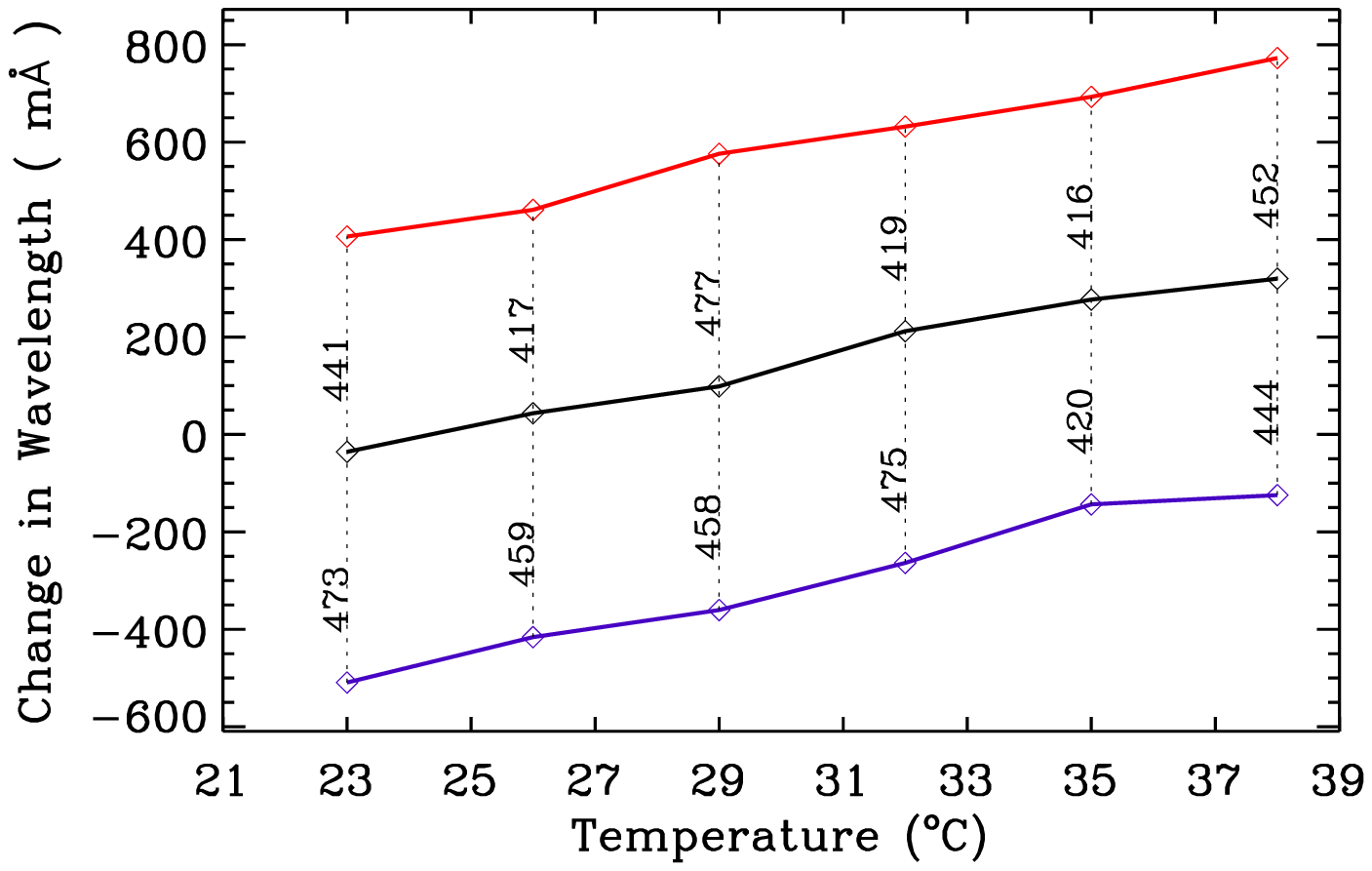}
\hspace {0.1cm}
\includegraphics[width=0.48\textwidth]{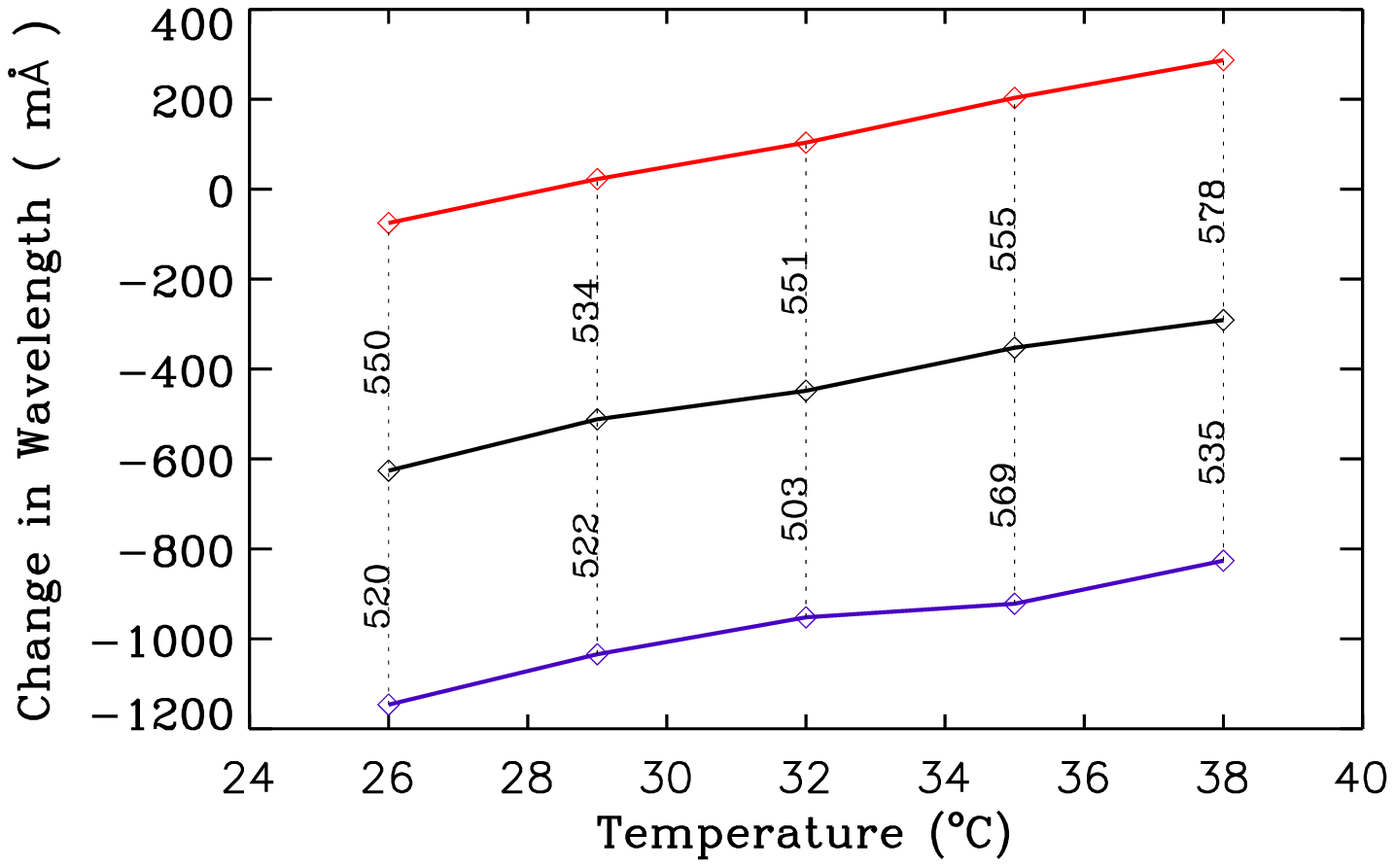}
\caption{Left and right panels show the temperature-sensitivity of FP$_2$ at 6173~\text{\AA} and 8542~\text{\AA}, respectively. Blue, black, and red continuous lines joining the diamonds represent the shift in wavelength with temperature for -3~kV, 0~V, and +3~kV, respectively. The values between the curves indicate the change in wavelength (in m\text{\AA}) corresponding to the change in voltage at a given temperature. The average slope is around 25 m\text{\AA}$^\circ$C$^{-1}$.}
\label{tempSens}
\end{center}
\end{figure}         

Figure~4 shows the results of these measurements for etalon FP$_2$. We plotted the shift in wavelength of the transmission in function of temeperature, close to the spectral lines of interest centered on 6173~\text{\AA} (left) and 8542~\text{\AA} (right). Blue, black, and red continuous lines joining the diamonds represent the shift in wavelength from line center for -3~kV, 0~V, and +3~kV, respectively. An increase in the temperature moves the transmission peaks to the higher wavelength side, indicating an increase in the refractive index of the LN etalon. 

From Figure~\ref{tempSens}, it is clear that the optimum temperature of FP$_2$ is 35\degC. At this temperature, the transmission channel of FP$_2$ can scan both line profiles optimally with the application of the voltage in the range of $\pm$3~kV. Similarly, the temperature tuning of FP$_1$ shows that the etalon kept at any temperature between 27\degC~and 33\degC~would allow us to scan the full line profile of both lines by applying voltages in the range of $\pm$3~kV.

\subsection{Voltage Tuning of the LN Etalons}

In this section, we explore how the voltage tunability of the etalons can be used to scan the two line profiles while keeping the etalons at an optimum temperature. As mentioned earlier, voltage tuning of the etalons is widely used to obtain high-cadence, multiwavelength observations. A linear high-voltage power supply with zero-crossing (from  Applied Kilovolts, UK, now a part of Exelis Inc., USA) was used to obtain an output of $\pm$5~kV for a computer-controlled input of $\pm$10~V in direct current (DC).  The details of the power supply unit and the control electronics can be found in \inlinecite{Shibu2010}. The etalons are operated by restricting the voltage between $\pm$3~kV and the rate of change of the voltage across the etalons is maintained at less than 1.5~kVs$^{-1}$. These values are restricted according to the CSIRO advice for safe operation of the etalons and are taken into account in the control software.

\begin{figure*}[!t]
\begin{center}
\includegraphics[width=0.320\textwidth, angle=90]{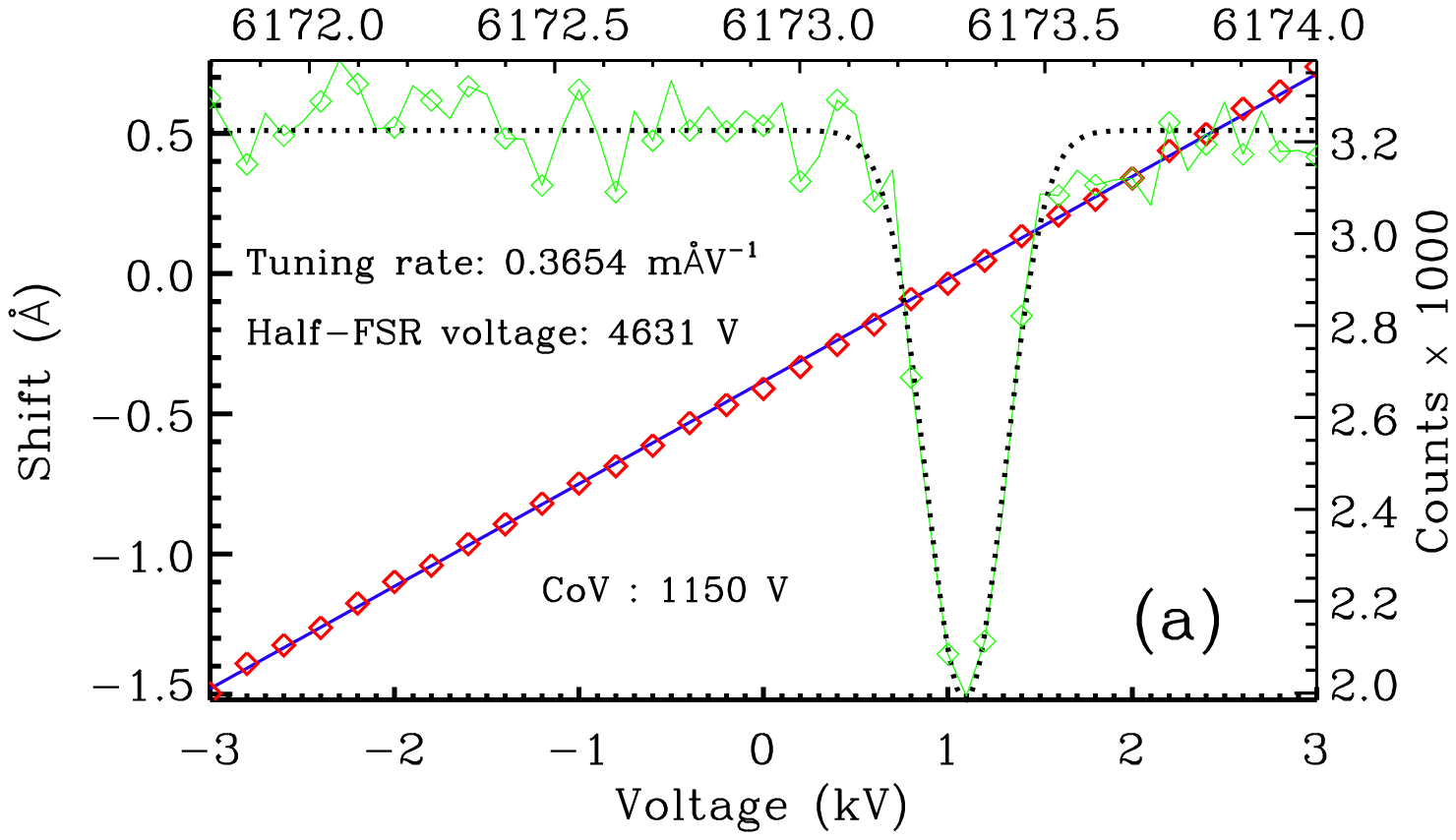}
\hspace{1mm}
\includegraphics[width=0.320\textwidth, angle=90]{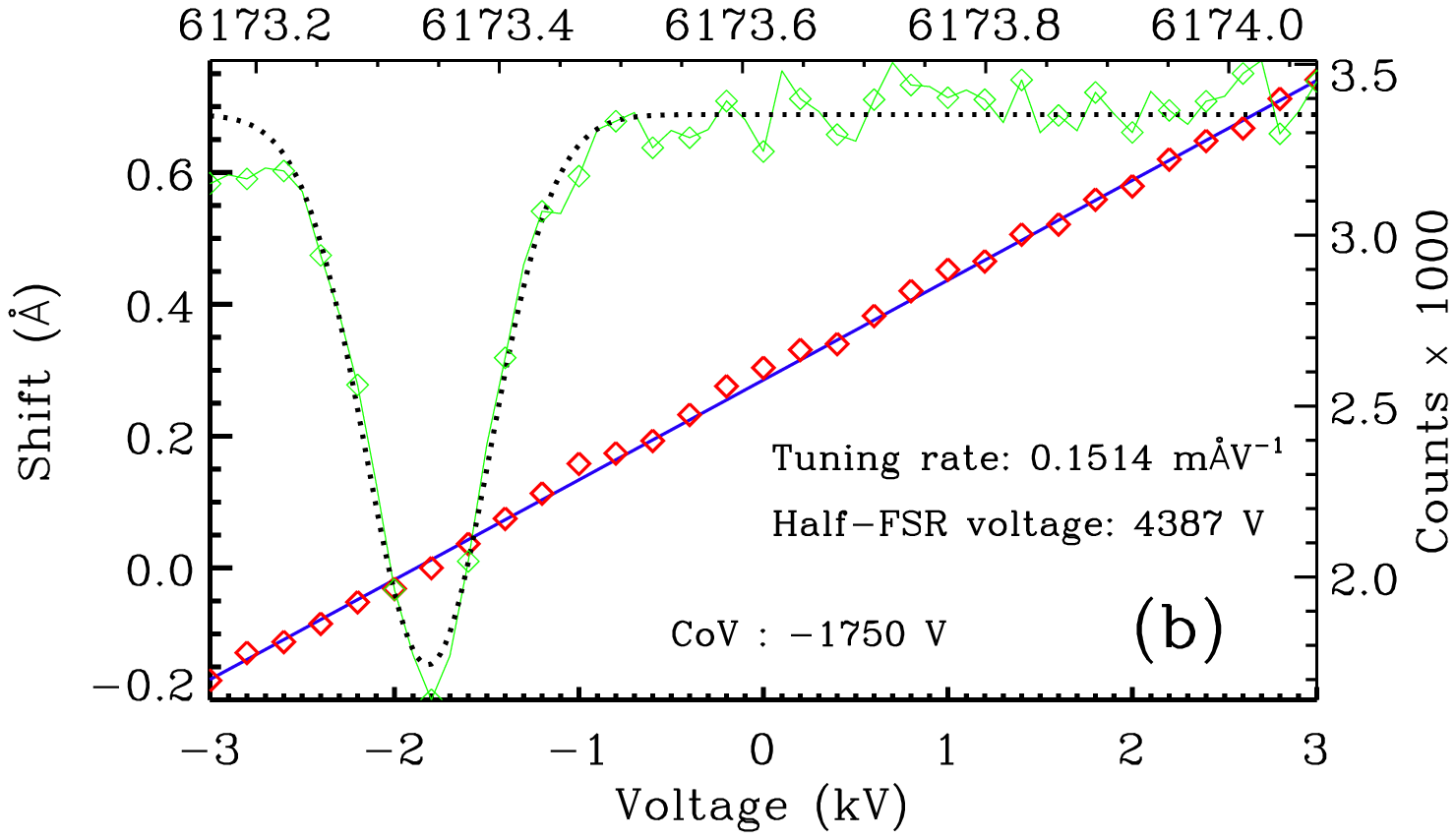}\\
\includegraphics[width=0.320\textwidth, angle=90]{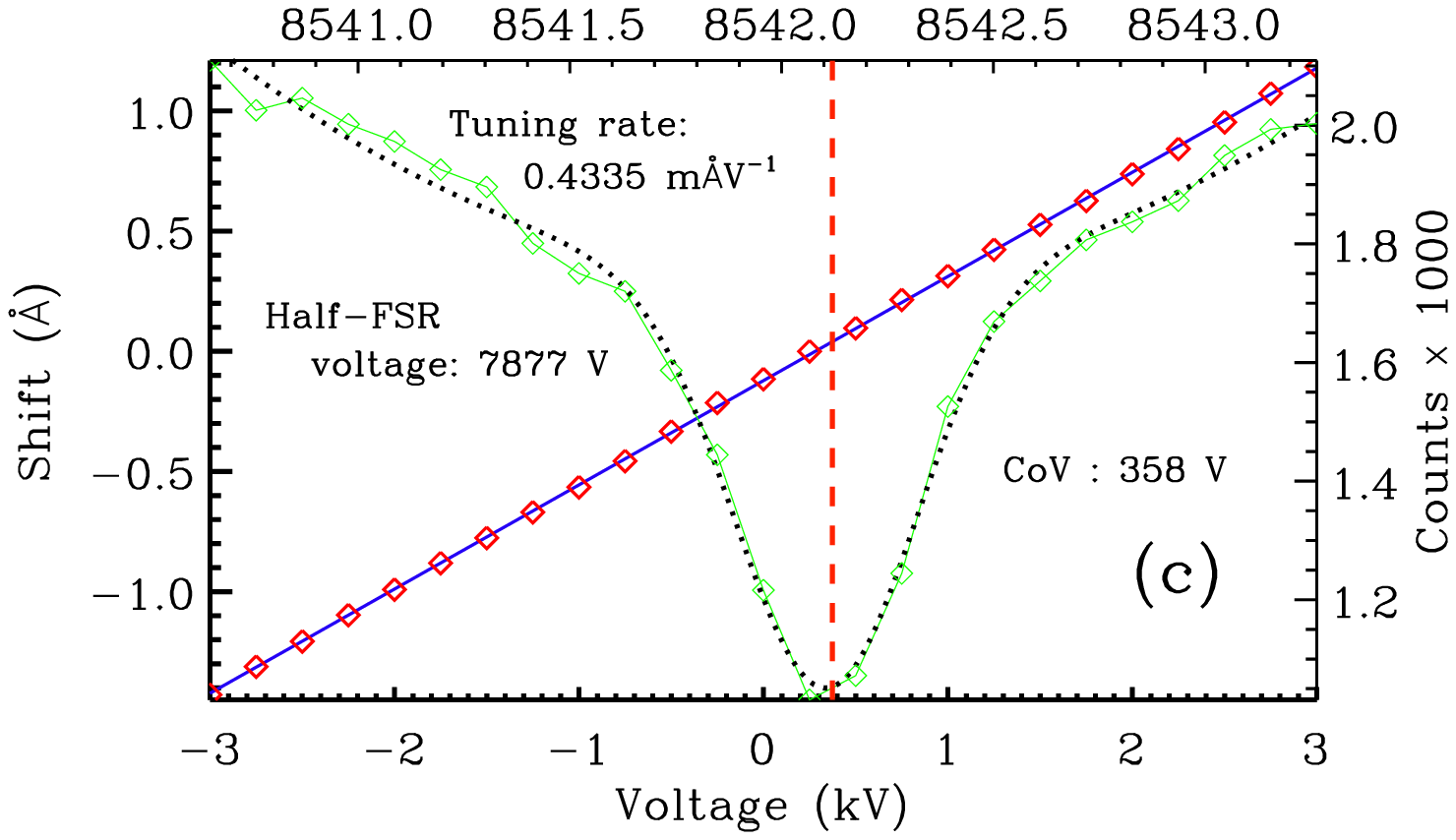}
\hspace{1mm}
\includegraphics[width=0.320\textwidth, angle=90]{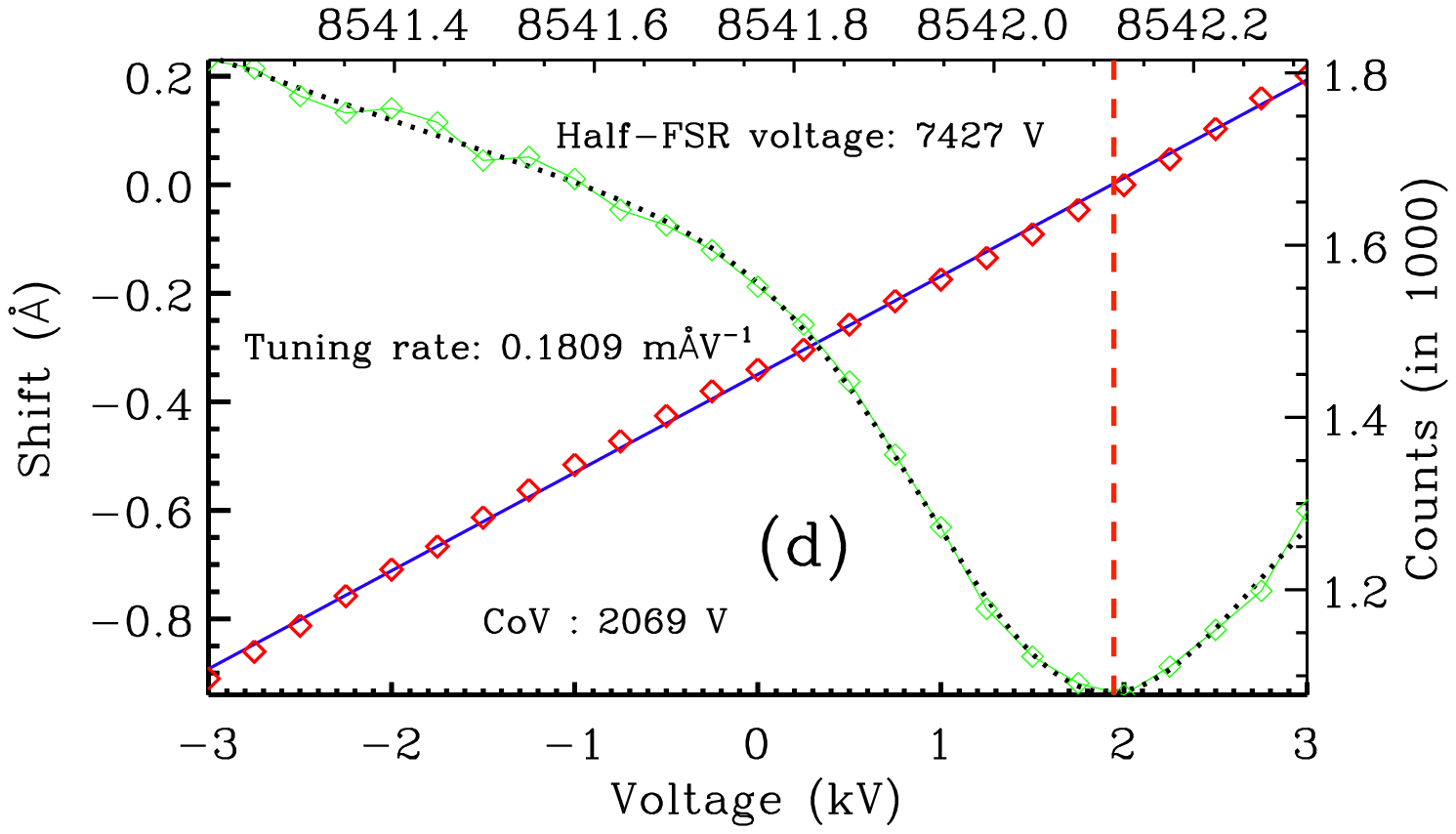}
\caption{Voltage tuning of the individual etalons at the lines of interest: (a) and (b) for FP$_1$ and FP$_2$ at 6173~\text{\AA}, respectively; (c) and (d) for 8542~\text{\AA} line, respectively. The panels show (i) the wavelength shift from the line center as a function of voltage with a blue continuous line and red diamonds and (ii) scanning of the spectral line with the application of the voltage to the individual etalons, which is shown as a green continous line with diamonds corresponding to the data counts vs. voltage. A Gaussian fit of this curve is shown as a black dotted line. The red vertical dashed line represents the line center. The corresponding wavelength axis is shown at the top of each panel. It is evident that for 6173~\text{\AA}, FP$_2$ has to be maintained at a temperature lower than 35{\degC},~whereas for 8542~\text{\AA}, a higher temperature ($\geq$35\degC) is required.} 
\label{voltage}
\end{center}
\end{figure*}
Figures 5a--d show the results obtained from the voltage tuning of both etalons for the two spectral lines by keeping the etalons FP$_1$ and FP$_2$ at a temperature of 30\degC~and 35\degC, respectively. The measurements were carried out by recording the transmission channel spectra for voltages from +3~kV to -3~kV with a step size of 100~V. The red diamonds represent the measured wavelength shift from $\lambda_0$ for the applied voltages (x-axis). The straight continuous line fitting the measurements allows us to estimate the voltage-tuning rate and the half-FSR voltage. The crossover voltage (CoV), {\it i.e.} the voltage at which the transmission channel crosses the center of the spectral line, is obtained from the green curve shown in Figure~5.

The voltage-tuning rate obtained from these measurements is indicated in the plots in units of \mAV along with the required half-FSR voltages. The estimated tuning rate ($\delta\lambda/V$) allows us to compute the unclamped electro-optic coefficient ($r_{13}$) of the LN crystals using Equation~(3). The mean value of $r_{13}$ for FP$_1$ and FP$_2$ at 6173~\text{\AA} is estimated as 5.08$\pm$0.06~\pmV and 5.30$\pm$0.15~\pmV~respectively, while the mean value of the $r_{13}$ coefficient at 8542~\text{\AA} is estimated to be 4.37~\pmV and 4.64$\pm$0.06~\pmV for FP$_1$ and FP$_2$, respectively. The $r_{13}$ coefficients estimated from voltage-tuning rates obtained at different temperatures were averaged to obtain the mean value of the coefficient. Different parameters obtained from temperature and voltage tuning of the etalons are shown in Table~3. The observed difference (244~V at 6173~\text{\AA} and 450~V at 8542~\text{\AA}) in the half-FSR voltages of FP$_1$ and FP$_2$ at a given wavelength might be due to the variation in the $r_{13}$ coefficients of FP$_1$ and FP$_2$ introduced during the crystal growth process. 

In this regard, we note that the values of the $r_{13}$ coefficients reported by various authors differ considerably. \inlinecite{ShibuAO} estimated the $r_{13}$ coefficient of a {\it z-}cut LN crystal to be 6.28~\pmV at 6122~\text{\AA}. This value differs from those obtained by other researchers. For example, \inlinecite{Turner} and \inlinecite{Onuki} reported a value of  8.6~\pmV and 10.9~\pmV measured at 6328~\text{\AA}, respectively. Similarly, \inlinecite{Burton1987} reported a value of 6.14~\pmV at 5800~\text{\AA}. Most of these values correspond to different wavelengths and, hence, it would be interesting to study the wavelength dependence of $r_{13}$.

As evident from Figure~5, the half-FSR voltages required to tune the etalons anywhere between the FSR are higher than the safe voltage that can be applied to the crystal. This makes it difficult to tune the etalons by applying a voltage to scan the entire line profile of Ca {\sc ii}~8542~\text{\AA}. It should be possible to  specify the central wavelength position of one of the channels to be exactly in the line center when fabricating the etalons. However, using the same etalon for multiwavelength observations is very difficult. Furthermore, for a broad spectral line like Ca~{\sc ii}~8542~\text{\AA}, the restriction in the applied voltage allows us to scan only a part of the line profile. The other possibility of tuning the etalons by varying their temperatures is also constrained by the recommended maximum operating temperature of 45\degC. We have not explored the tilt tuning of the etalons. Moreover, as the reflective coatings of the etalons are optimized for normal incidence, reflection losses are expected at larger tilt and, thus, the effective finesse of the etalons reduces. Therefore, tilt tuning is not a good option.

\begin{table}[t]
\caption{Parameters obtained by temperature and voltage tuning of the etalons}
\begin{tabular}{c|c|c|c|c|c|c|c}
\hline
Wave-       & Fabry-   & Optimum      &   VT$^\bot$ & CoV$^\top$ & V$_{\frac{1}{2}FSR}$& Tuning        & $r_{13}$  \\
length      & Perot    & Temperature  & (\mAV)   &  (V)       & (V)                 & ratio         & (\pmV)    \\
\hline
6173~\AA    & FP$_1$   & 30$^\circ$C  &  0.36541    &   $+1150$  & 4631                &  2.413        & 5.08      \\
      	    & FP$_2$   & 35$^\circ$C  &  0.15147    &   $-1750$  & 4387                &               & 5.30      \\
\hline
8542~\AA    & FP$_1$   & 30$^\circ$C  & 0.43353     &   $+0358$  & 7877                &   2.396       & 4.37      \\
      	    & FP$_2$   & 35$^\circ$C  &  0.18094    &   $+2069$  & 7427                &               & 4.64      \\
\hline
\end{tabular}
$\bot$\;:\;Voltage tuning rate, \quad $\top$\;:\;Crossover Voltage, \quad $V_{\frac{1}{2}FSR}$\;:\;Half-FSR voltage\\
\end{table}

\subsection{Voltage Tuning of the LN Etalons in Tandem}  

The etalons were tuned in tandem using the values obtained after temperature and voltage tuning of the individual etalons. The temperature for FP$_1$ and FP$_2$ was fixed at 30$^\circ$C and 35$^\circ$C, respectively. The voltage-tuning values indicated in Figure~5 and listed in Table~3 were used to tune the etalons in tandem to scan the line profiles. Crossover voltages of the individual etalons were used to obtain the transmission profile of the FPs in tandem at the center of the lines of interest, and the voltage-tuning rates were used to set the transmission peaks of both etalons to the same wavelength. For example, the ratio of the voltage-tuning rates of FP$_1$ and FP$_2$ at 6173~\text{\AA} and 8542~\text{\AA} are 2.413 and 2.396, respectively. This means that FP$_2$ requires 2.413~V for each 1~V change of FP$_1$ to shift the transmission channel to the same wavelength position. Similarly, it requires 2.396~V for observations at 8542~\text{\AA}. The scanning of the spectral line in any desired wavelength step can be carried out by a control program provided the voltage tuning rates and crossover voltages are known.

\begin{figure*}[b]
\begin{center}
\includegraphics[width=0.165\textwidth, angle = 90]{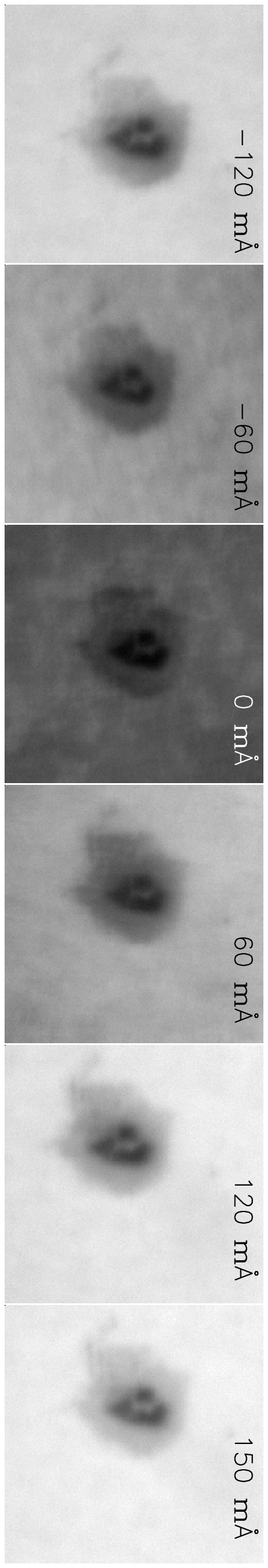}
\includegraphics[width=0.165\textwidth, angle = 90]{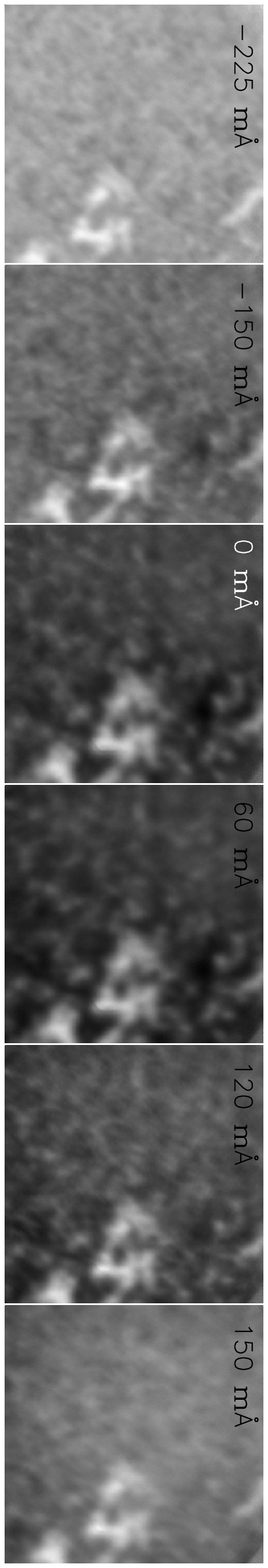}
\caption{Images obtained at 6173~\text{\AA} (top) and 8542~\text{\AA} (bottom), at the indicated wavelength positions during the corresponding line scan. The FOV for the top panels is 150~arcsec$^2$ and that for the bottom panels is 300~arcsec$^2$.}
\label{scanImages}
\end{center}
\end{figure*} 

\begin{figure*}[t]
\begin{center}
\includegraphics[width=0.45\textwidth]{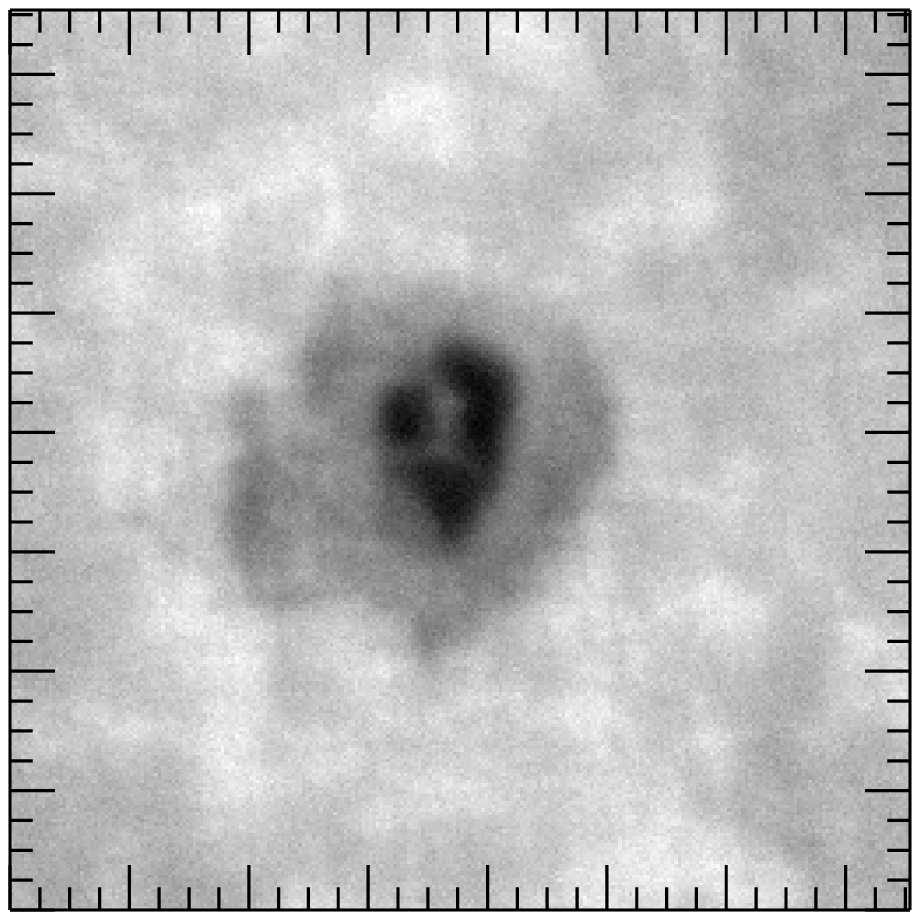}
\includegraphics[width=0.45\textwidth]{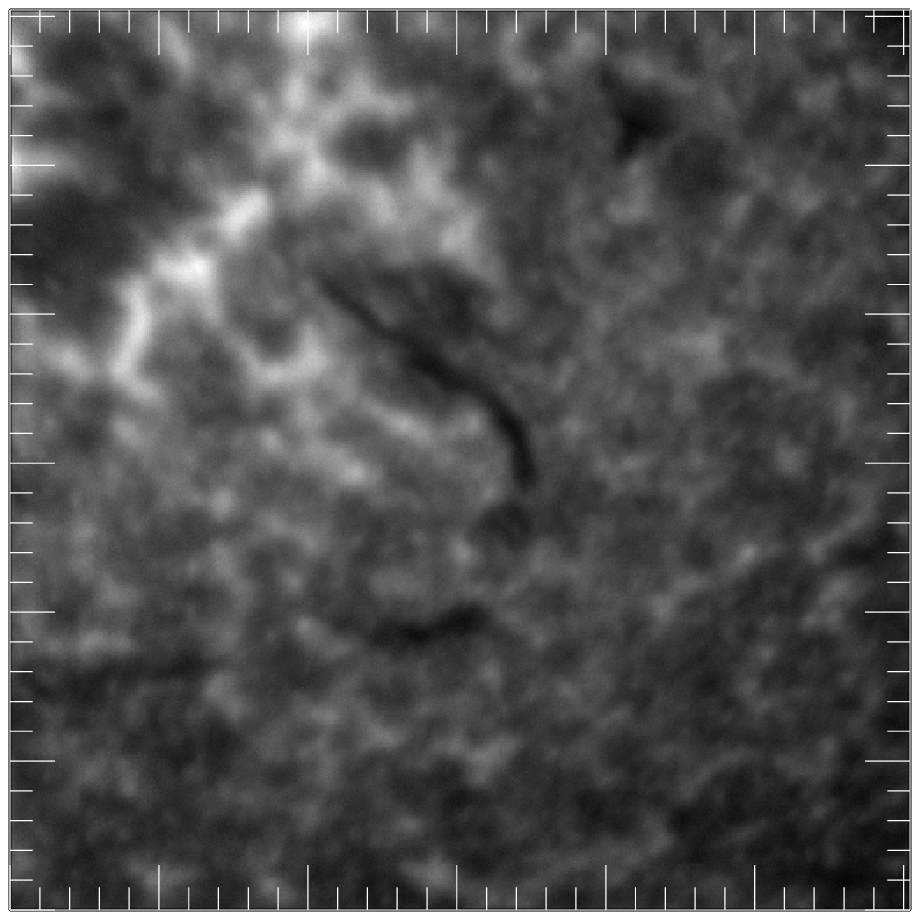}
\caption{Images obtained by tuning the filter on the line center of (a) 6173~\text{\AA} 
and (b)  8542~\text{\AA}. The FOV of (a) is 150~arcsec$^2$ and of (b) 300~arcsec$^2$.}
\label{LCimages}
\end{center}
\end{figure*}

\section{Preliminary Observations with the Narrow-Band Imager}

After carrying out the calibration runs using the spectrograph, the collimated beam from the FPs was diverted toward the imaging system using a beam-steering mirror (BSM) as described in Section 3. A filter wheel with two interference filters centered on 6173~\text{\AA} (FWHM = 3.0~\text{\AA}) and 8542~\text{\AA} (FWHM = 3.5~\text{\AA}) was employed to select the desired wavelength (channel) for observations. A series of images was obtained by tuning the etalons in tandem across different wavelength positions on both spectral lines. Images acquired at different wavelength positions are shown in Figure 6. Figure~7 shows a magnified view of the images taken in the line center at 6173~\text{\AA} and 8542~\text{\AA}, respectively. Observations in 6173~\text{\AA} were obtained on 3 April 2013, the sunspot observed was part of the active region (AR) NOAA 11711 located at S17E24 on the solar disk. Observations in 8542~\text{\AA} were obtained on 8 March 2013, the filament that is seen in the image was located on the disk at S35E25 near AR 11689. 

For the 6173~\text{\AA} line, the images were obtained from -150~m\text{\AA} to +210~m\text{\AA} from line center, with a step of 30~m\text{\AA}, whereas the line profile at 8542~\text{\AA} was scanned from -225~m\text{\AA} to +150~m\text{\AA} in a step of 75~m\text{\AA}. It is noteworthy here that the quality of these images is affected by the poor seeing conditions at the site\footnote{The 15~cm Coud\'{e} telescope used for calibration and acquiring test images is located next to a laboratory building. However, the narrow-band imager will be integrated with a 50 cm telescope MAST at the island observatory, where the seeing conditions are relatively better.}. Using similar observations obtained for the quiet Sun, we retrieved the profiles for both spectral lines (Figure 8). The mean intensity of the quiet Sun after flat-fielding and dark-current correction were used in retrieving these line profiles. As evident, because of the limited tunability of the filter, we were unable to retrieve the complete line profile of the 8542~\text{\AA} line. The exposure times used to obtain each filtergram in 6173~\text{\AA} and 8542~\text{\AA} are 120~ms and 900~ms, respectively. However, with a 50 cm telescope the exposure times are expected to be of the order of 50~ms and 300~ms, respectively, to build the signal to a level of 80$\%$ of the full-well capacity ({\it i.e.} 16000 e$^{-}$) of the CCD.

\begin{figure*}[!h]
\begin{center}
\includegraphics[width=0.47\textwidth]{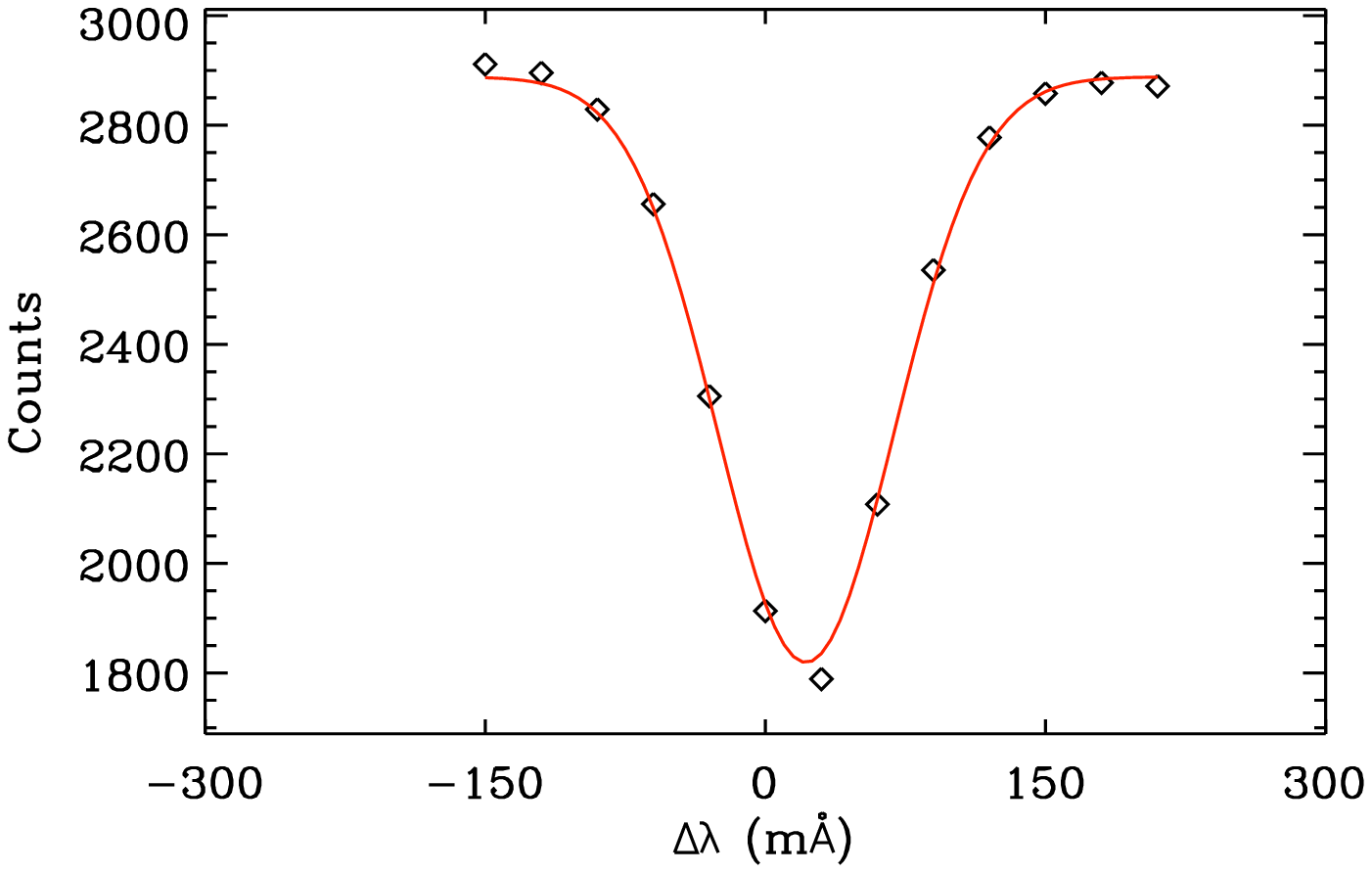}
\includegraphics[width=0.47\textwidth]{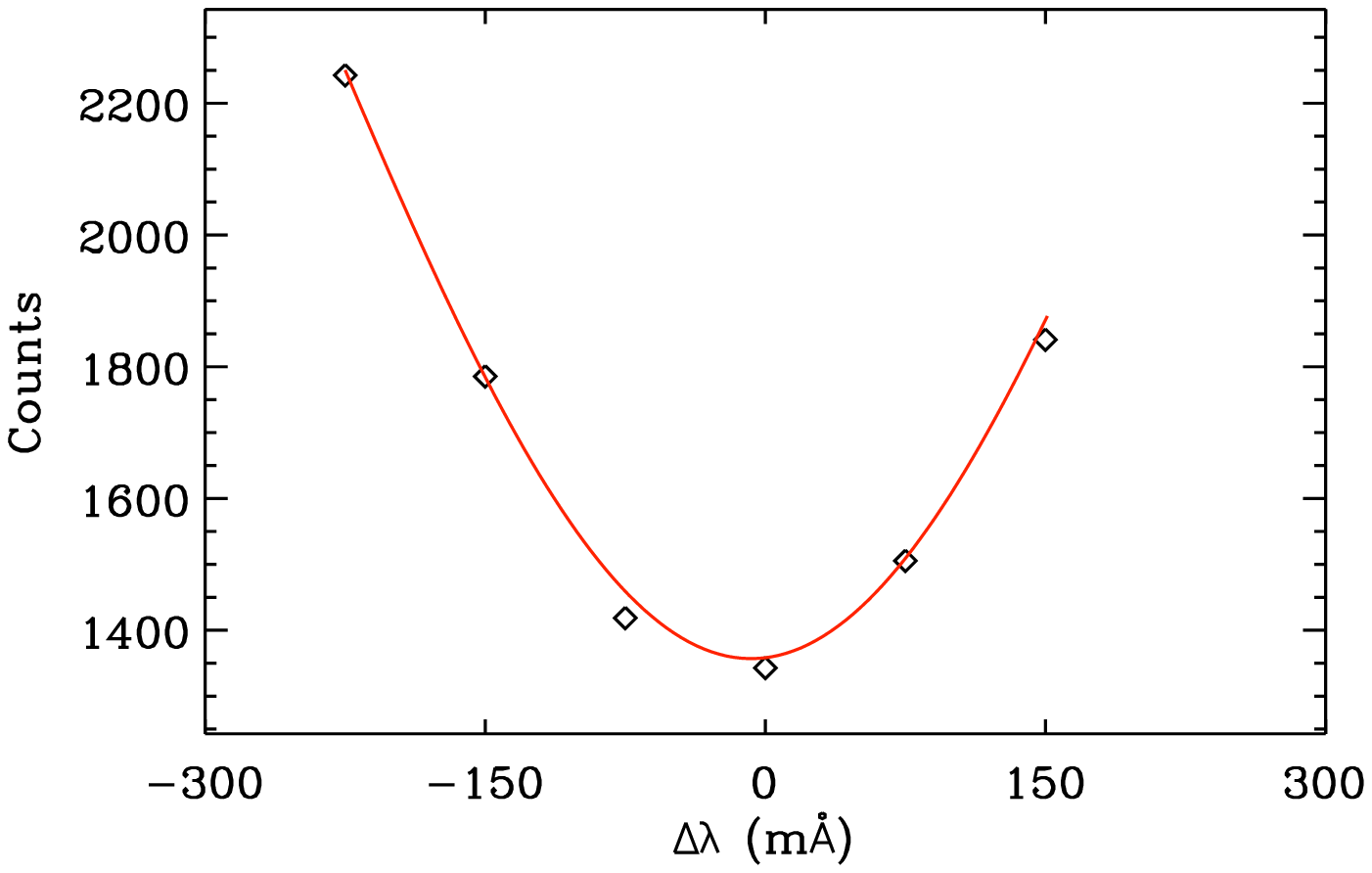}
\caption{Retrieved line profiles at 6173~\text{\AA} (left) and 8542~\text{\AA} (right) using quiet-Sun observations. The line profile at 6173~\text{\AA} is scanned at 13 wavelength positions from -150 m\text{\AA} to +210 m\text{\AA} with $\delta\lambda$ = 30 m\text{\AA}. The line profile at 8542~\text{\AA} is scanned only in the core at five wavelength positions from -225~m\text{\AA} to +150~m\text{\AA} with $\delta\lambda$ = 75~m\text{\AA}.}
\label{LineRetr}
\end{center}
\end{figure*}

\section{Conclusions}
We demonstrated the use of two lithium niobate etalons (FP$_1$ and FP$_2$) in tandem for quasi-simultaneous narrow-band observations in Fe~{\sc i} (6173~\text{\AA}) and Ca~{\sc ii}~(8542~\text{\AA}). Temperature and voltage-tuning calibrations for both etalons were carried out using a Littrow spectrograph. The temperatures of FP$_1$ and FP$_2$ were fixed at 30$^\circ$C and 35$^\circ$C, respectively. The ratio of voltage-tuning rates of FP$_1$ and FP$_2$ at 6173~\text{\AA} and 8542~\text{\AA} were found to be 2.413 and 2.396, respectively. The computed values of the $r_{13}$ coefficients for the lithium niobate crystals were found to be of the order of 5.2~\pmV and 4.5~\pmV at 6173~\text{\AA} and 8542~\text{\AA}, respectively.

A Coud\'{e} telescope with a clear aperture of 15~cm was used to obtain the test images in both lines. The line profiles of these two lines of interest were retrieved using quiet-Sun observations. However, given the limitations, the system can scan the 8542~\text{\AA} line only partially. This system will be integrated with the upcoming solar telescope, MAST, in about mid-2014. We did not perform a rigorous ghost analysis here. We plan to use a prefilter in between the two etalons to deal with the ghost images after integrating the instrument with MAST. The ghost-image analysis will be carried out for the system, and if required, the etalons will be tilted to minimize the problem by avoiding any significant broadening of the filter profile \cite{Valentine2011}.

\begin{acks}
We acknowledge the work done by S. K. Gupta to develop the control electronics for the high-voltage power-supply unit and temperature controller. We also acknowledge the work done by M. Sardava for the design and fabrication of the temperature-controlled enclosures for the FPs and the prefilter wheel, is also acknowledged. We thank the anonymous referee for his/her valuable suggestions and comments that improved the quality of the article.
\end{acks}     

{}


\begin{thebibliography}{}

\bibitem[\protect\citeauthoryear{{Barthol} {\it et~al.}}{2011}]{sunrise}
Barthol, P., Gandorfer, A., Solanki, S.K., Sch{\"u}ssler, M., Chares, B., Curdt, W., {\it et al.}.: 2011, {\it Solar Phys.} {\bf 268}, 1. 

\bibitem[\protect\citeauthoryear{{Beck}, {Rezaei} and {Fabbian}}{2011}]{Beck2011}
Beck, C., Rezaei, R., Fabbian, D.: 2011, {\it Astron. Astroph.} {\bf 535}, A129. 

\bibitem[\protect\citeauthoryear{{Bello Gonz{\'a}lez} and {Kneer}}{2008}]{GPI2008}
Bello Gonz{\'a}lez, N.~and Kneer, F.: 2008, {\it Astron. Astroph.} {\bf 480}, 265. 

\bibitem[\protect\citeauthoryear{{Bonaccini and {Smartt}}}{1988}]{Bonaccini}
{Bonaccini}, D., {Smartt}, R., N.: 1988, \AO~{\bf 27}, 5095.

\bibitem[\protect\citeauthoryear{{Born and Wolf}}{1989}]{BornWolf}
{Born},~M., {Wolf},~E.: 1989, {\it Principles of Optics}, Pergamon Press.

\bibitem[\protect\citeauthoryear{{Borrero} {\it et~al.}}{2011}]{Borrero2011}
Borrero, J.M., Tomczyk, S., Kubo, M., Socas-Navarro, H., Schou, J., Couvidat, S., Bogart, R.: 2011, {\it Solar Phys.} {\bf 273}, 267.

\bibitem[\protect\citeauthoryear{{Burton}, {Leistner} and {Rust}}{1987}]{Burton1987}
{Burton}, C.~H., {Leistner}, A.~J., {Rust}, D.~M.: 1987, \AO~{\bf 26}, 2637.

\bibitem[\protect\citeauthoryear{{Cavallini}}{2006}]{IBIS2006}
{Cavallini}, F.: 2006, \solphys~{\bf 236}, 415.
 
\bibitem[\protect\citeauthoryear{{Desai}}{1996}]{Desai1996}
{Desai}, J.~N.: 1996, \BASI~{\bf 24}, 849.

\bibitem[\protect\citeauthoryear{{Delbouille}, {Roland} and {Neven}}{1973}]{BASS2000}
{Delbouille}, L., {Roland}, G., {Neven}, L.: 1973, Atlas photometrique du spectre solaire de  {$\lambda$} 3000 a {$\lambda$} 10 000,
Universite de Liege, Institut d' Astrophysique, Liege.
  

\bibitem[\protect\citeauthoryear{{Denis} {\it et~al.}}{2008}]{AMOS2008}
{Denis}, S., {Coucke}, P., {Gabriel}, E. {\it et~al.}: 2008, {\it Proc. SPIE} {\bf7012}, id. 701235.
 
\bibitem[\protect\citeauthoryear{{Denis} {\it et~al.}}{2010}]{AMOS2010}
{Denis}, S., {Coucke}, P., {Gabriel}, E. {\it et~al.}: 2010, {\it Proc. SPIE} {\bf7733}, id. 773335.
 
\bibitem[\protect\citeauthoryear{{Evans}}{1948}]{John}
{Evans}, J. W.: 1948, \josa~{\bf 30}, 229.

\bibitem[\protect\citeauthoryear{{Fahmy} {\it et~al.}}{2013}]{Fahmy}
{Fahmy}, S. {\it et~al.}: 2013 {\it Proc. 64 Int. Astron. Con.} {\bf IAC-13}, A3.5.2
 
\bibitem[\protect\citeauthoryear{{Gosain} {\it et~al.}}{2006}]{SVM2006}
{Gosain}, S., {Venkatakrishnan}, P., {Venugopalan}, K.: 2006, \jaa~{\bf 27}, 285.
 
\bibitem[\protect\citeauthoryear{{Ghatak} and {Thyagarajan}}{1989}]{Ghatak}
{Ghatak}, A., Thyagarajan, K.: 1989, {\it Optical electronics}, Cambridge University Press.
 
\bibitem[\protect\citeauthoryear{{Judge \it{et~al.}}}{2010}]{Judge2010}
Judge, P.G., Tritschler, A., Uitenbroek, H., Reardon, K., Cauzzi, G., de Wijn, A.: 2010, {\it Astrophys. J.} {\bf 710}, 1486. 
 
\bibitem[\protect\citeauthoryear{{Kentischer} {\it et~al.}}{1998}]{Kentischer1998}
Kentischer, T.J., Schmidt, W., Sigwarth, M., and Uexkuell, M.V.: 1998, {\it Astron. Astroph.} {\bf 340}, 569. 

\bibitem[\protect\citeauthoryear{{Kleint}, {Feller} and {Gisler}}{2011}]{Kleint2011}
{Kleint}, L., {Feller}, A., {Gisler}, D.: 2011, \aap~{\bf 529}, A78.
 
\bibitem[\protect\citeauthoryear{{Lagg} {\it et~al.}}{2004}]{MEinversion}
{Lagg}, A., {Woch}, J., {Krupp}, N., {Solanki}, S.~K.: 2004, \aap~{\bf 414}, 1109.

\bibitem [\protect\citeauthoryear{{Lyot}}{1933}]{Lyot}
{Lyot},~B.: 1993, {\it Compt Rend. Acad. Sci. Paris} {\bf 197}, 1593.
 
\bibitem[\protect\citeauthoryear{{Mart\'{\i}nez Pillet} {\it et~al.}}{2011}]{Valentine2011}
Mart{\'{\i}}nez Pillet, V., Del Toro Iniesta, J.C., {\'A}lvarez-Herrero, A., Domingo, V., Bonet, J.A., 
Gonz{\'a}lez Fern{\'a}ndez, {\it et~al.}: 2011, {\it Solar Phys.} {\bf 268}, 57. 
 
\bibitem[\protect\citeauthoryear{{Mathew} {\it et~al.}}{1998}]{Shibu1998}
{Mathew}, S.~K., {Bhatnagar}, A., {Prasad}, C.~D., {Ambastha}, A.: 1998, \aaps~{\bf 133}, 285.
 
\bibitem[\protect\citeauthoryear{{Mathew} {\it et~al.}}{2001}]{Shibu_so}
{Mathew}, S.~K., {Solanki}, S. K., and VIM team.: 2001, in S.~C.~Tripathy, P.~Venkatakrishnan (ed.) {\it Probing the Sun with high resolution},  Narosa Pub. House, 213.

\bibitem[\protect\citeauthoryear{{Mathew}}{2003}]{ShibuAO}
{Mathew}, S.~K.: 2003, {\it Appl. Opt.}~{\bf 42}, 3580.

\bibitem[\protect\citeauthoryear{{Mathew}}{2009}]{ShibuMAST}
{Mathew}, S.~K.: 2009, in S.V. Berdyugina, S.K., Nagendra, R. Ramelli (ed.) {\it Solar Polarization 5: In Honor of Jan Stenflo} {\bf 405}, 461. 
 
\bibitem[\protect\citeauthoryear{{Mathew} and {Gupta}}{2010}]{Shibu2010}
{Mathew}, S.~K., {Gupta}, S.~K.: 2010, {\it Tech. Rep.} {\bf PRL-TN-2010-97}, Phys. Res. Lab., Ahmedabad, India. 

\bibitem[\protect\citeauthoryear{{Netterfield} {\it et~al.}}{1997}]{Netterfield}
{Netterfield},~R.~P., {Freund},~C.~H, {Seckold}~J.~A., {Walsh},~C.~J.: 1997, {\it Appl. Opt.} {\bf 36}, 4556.

\bibitem[\protect\citeauthoryear{{Onuki}, {Uchida} and {Saku}}{1972}]{Onuki}
{Onuki},~K., {Uchida},~N., {Saku},~T.: 1972, \josa~{\bf 62}, 1030.

\bibitem[\protect\citeauthoryear{{Prasad Choudhary} and {Gosain}}{2002}]{Gosain2002}
{Prasad Choudhary}, D., {Gosain}, S.: 2002, {\it Exp. Astron.} {\bf 13}, 153. 
 
\bibitem[\protect\citeauthoryear {Prasad Choudhary {\it et al.}}{1998}]{Shibu1997}
{Prasad Choudhary}, D., {Mathew}, S., {Bhatnagar}, A., {Ambastha}, A.: 1998, {\it Exp. 
Astron.} {\bf 8}, 125. 

\bibitem[\protect\citeauthoryear{{Rust} {\it et~al.}}{1996}]{Rust}
Rust, D.M., Murphy, G., Strohbehn, K., Keller, C.U.: 1996, {\it Solar Phys.} {\bf 164}, 403. 
 
\bibitem[\protect\citeauthoryear{{Scharmer}}{2006}]{Scharmer2006}
{Scharmer}, G.~B.: 2006, \aap~{\bf 447}, 1111.

\bibitem[\protect\citeauthoryear{{Scharmer} {\it et~al.}}{2008}]{CRISPSc2}
Scharmer, G.~B., Narayan, G., Hillberg, T., de la Cruz Rodr{\'{\i}}guez, J., L{\"o}fdahl, 
M.G., Kiselman, D., {\it et all.}: 2008, {\it Astrophys. J.} {\bf 689}, L69. 

\bibitem[\protect\citeauthoryear{{Schuhle} {\it et~al.}}{2007}]{Schuhle}
Sch{\"u}hle, U., Mathew, S.K., Wedemeier, M., Hartwig, H., Ballesteros, E., Martinez 
Pillet, V., Solanki, S.K.: 2007, {\it ESA Special Publication} {\bf 641}, 82.

\bibitem[\protect\citeauthoryear{{Stix}}{1991}]{Stix}
{Stix},~M.: 1991, {\it The Sun: An Introduction}, Springer-Verlag. 
 
\bibitem[\protect\citeauthoryear{{Tritschler} {\it et~al.}}{2002}]{Tritschler2002}
{Tritschler}, A., {Schmidt}, W.: {Langhans}, K., {Kentischer}, T.: 2002, \solphys, {\bf 211}, 17.

\bibitem[\protect\citeauthoryear{{Turner}}{1966}]{Turner}
{Turner},~E.~H.: 1966, {\it Appl. Phys. Lett.} {\bf 8}, 303.

\bibitem[\protect\citeauthoryear{{Wiegelmann} {\it et~al.}}{2010}]{wiegelmann}
Wiegelmann, T., Solanki, S.K., Borrero, J.M., Mart{\'{\i}}nez Pillet, V., del Toro 
Iniesta, J.C., Domingo, V., {\it et al.}: 2010, {\it Astrophys. J.} {\bf 723}, L185. 

\bibitem[\protect\citeauthoryear{{Yariv}}{1984}]{Yariv}
{Yariv},~A.: 1984, {\it Optical electronics}, Holt McDougal, Third Ed.

\bibitem[\protect\citeauthoryear{{Zirin}}{1995}]{Zirin}
{Zirin},~H.: 1995, \solphys~{\bf 159}, 203.

\end{thebibliography}
\end{document}